\documentclass[twocolumn,tighten]{aastex62}
\usepackage{natbib}
\usepackage{amsmath,amstext,amssymb}
\usepackage{listings}
\usepackage[normalem]{ulem}
\usepackage[section]{placeins}
\usepackage{color}

\usepackage{hyperref}
\usepackage{graphicx}
\usepackage[percent]{overpic}

\newcommand{\comments}[1]{} 

\shorttitle{Evolution and Mergers of LISA Massive BH Binaries in Disk Galaxy Mergers}
\shortauthors{F.~M.~Khan et al.}

\setwatermarkfontsize{1.6in}

\begin{document}

\title{Dynamical Evolution and Merger Time-scales of LISA Massive Black Hole Binaries in Disk Galaxy Mergers}

\correspondingauthor{Fazeel~M.~Khan}
\email{khanfazeel.ist@gmail.com}

\author[0000-0002-5707-4268]{Fazeel~M.~Khan}
\affiliation{Department of Space Science, Institute of Space Technology, P.O. Box 2750, Islamabad 44000, Pakistan}

\author[0000-0002-1786-963X]{Pedro~R.~Capelo}
\affiliation{Center for Theoretical Astrophysics and Cosmology, Institute for Computational Science, University of Zurich, Winterthurerstrasse 190, CH-8057 Z\"{u}rich, Switzerland}

\author[0000-0002-7078-2074]{Lucio~Mayer}
\affiliation{Center for Theoretical Astrophysics and Cosmology, Institute for Computational Science, University of Zurich, Winterthurerstrasse 190, CH-8057 Z\"{u}rich, Switzerland}

\author[0000-0003-4176-152X]{Peter~Berczik}
\affiliation{National Astronomical Observatories of China and Key Laboratory for Computational Astrophysics, Chinese Academy of Sciences, 20A Datun Rd, Chaoyang District, 100012 Beijing, People's Republic of China}
\affiliation{Astronomisches Rechen-Institut, Zentrum f\"{u}r Astronomie, University of Heidelberg, M\"{o}nchhofstrasse 12-14, DE-69120 Heidelberg, Germany}
\affiliation{Main Astronomical Observatory, National Academy of Sciences of Ukraine, 27 Akademika Zabolotnoho St., UA-03680 Kyiv, Ukraine}


\begin{abstract}
The Laser Interferometer Space Antenna (LISA) will detect gravitational-wave (GW) signals from merging supermassive black holes (BHs) with masses below $10^7$~M$_{\odot}$. It is thus of paramount importance to understand the orbital dynamics of these relatively light central BHs, which typically reside in disc-dominated galaxies, in order to produce reliable forecasts of merger rates. To this aim, realistic simulations probing BH dynamics in unequal-mass disc galaxy mergers, into and beyond the binary hardening stage, are performed by combining smooth particle hydrodynamics and direct $N$-body codes. The structural properties and orbits of the galaxies are chosen to be consistent with the results of galaxy formation simulations. Stellar and dark matter distributions are triaxial down to the central 100 pc of merger remnant. In all cases, a BH binary forms and hardens on time-scales of at most 100~Myr, coalescing on another few hundred Myr time-scale, depending on the characteristic density and orbital eccentricity. Overall, the sinking of the BH binary takes no more than $\sim$0.5~Gyr after the merger of the two galaxies is completed, but can be much faster for very plunging orbits. Comparing with previous numerical simulations following the decay of BHs in massive early-type galaxies at $z \sim 3$, we confirm that the characteristic density is the most crucial parameter determining the overall BH merging time-scale, despite the structural diversity of the host galaxies. Our results lay down the basis for robust forecasts of LISA event rates in the case of merging BHs.
\end{abstract}



\section{Introduction}\label{sec:Introduction}

Central supermassive black holes (BHs), with masses in the range $10^5$--$10^{10}$~M$_{\odot}$, are ubiquitous in galaxies of a wide range of masses, from dwarf galaxies to the most massive early-type galaxies \citep{Kormendy_Richstone_1995,Ferrarese_Ford_2005,Mezcua_et_al_2018}. Their masses correlate well with various properties of their host galaxies such as the mass and velocity dispersion of the stellar spheroid, their total stellar mass etc. \citep[][]{Gultekin_et_al_2009,Kormendy_Ho_2013,McConnell_Ma_2013,Graham_2016}, suggesting a tight link between the growth of BHs and that of their hosts. In hierarchical structure formation, within the concordance cosmological model, $\Lambda$-CDM, mergers between galaxies drive their mass assembly over time. The merger rate of galaxies increases fairly steeply with redshift, although the exact scaling relation is debated in both theoretical modelling and empirical determination via observations \citep[][]{Fakhouri_et_al_2010}. During mergers the expectation is that the central BHs will pair and bind into a binary eventually coalescing and becoming the loudest type of gravitational wave (GW) source once their separation shrinks to milliparsec \citep[][]{Begelman_et_al_1980,Colpi_Dotti_2011,Mayer_2013}. The Laser Interferometer Space Antenna (LISA) will be able to detect GWs emitted during the inspiral phase of BHs up to $z \sim 10$, and its frequency coverage is particularly favourable to detect coalescing BH binaries with masses in the range $10^3$--$10^7$~M$_{\odot}$. While for the low-mass end of such BHs, called intermediate-mass BHs, both observational (see, e.g. \citealt{Mezcua_2017} for a review) and numerical \citep[e.g.][]{Bellovary_et_al_2018,Tamfal_et_al_2018} studies are still scarce, evidence for BHs in the mass range $10^5$--$10^6$~M$_{\odot}$ is solid, coming from both observations of kinematics of galactic nuclei via stellar velocity fields, masers, and detections via X-ray, ultraviolet, etc. when the BH is active \citep[][]{Kormendy_Ho_2013}. The latter BHs reside at the centre of galactic bulges in present-day spiral galaxies. The processes that govern the evolution of the BH pair evolution across orders of magnitude in separation scale are diverse, from dynamical friction by the stellar, dark matter, and gaseous background \citep[][]{Callegari_et_al_2009}, to three-body encounters with incoming stars once the binary has become hard, at pc separations \citep[][]{Khan_et_al_2012b,Gualandris_Merritt_2012,Rantala_et_al_2017}, to torques induced by spiral density waves and other asymmetries when the BH binary is embedded in a mostly gaseous circumnuclear or circumbinary disc \citep[][]{Fiacconi_et_al_2013,Mayer_2013,Farris_et_al_2014,Ryan_MacFadyen_2017}.

In the last decade there has been considerable effort in modelling the orbital decay phases of massive BH pairs in galaxy mergers, using predominantly either numerical simulations that follow the BH binary to very small separations but capture only the gravitational dynamics of the stellar and dark matter components \citep[][]{Milosavljevic_Merritt_2001,Berczik_et_al_2006,Khan_et_al_2011}, or simulations the include the interaction with the gaseous interstellar medium (ISM) but normally cannot follow the decay process beyond pc scales \citep[][]{Escala_et_al_2005,Dotti_et_al_2006,Dotti_et_al_2007,Mayer_et_al_2007,Callegari_et_al_2009,Chapon_et_al_2013,SouzaLima_et_al_2017}. Achieving high enough resolution to model the hard-binary phase in simulations of galaxy mergers that include also hydrodynamics in the galaxy merger phase has been first attempted by \citet{Khan_et_al_2012a} but in a limited form. The merger time-scale was predicted in some cases by extrapolating the decay rate in the last phases of the simulation, typically obtaining long time-scales of a few to several Gyr \citep[][]{Khan_et_al_2011,Khan_et_al_2012b,Callegari_et_al_2009}, and in some cases finding even evidence for a possible stalling of binaries at pc separations \citep[][]{Chapon_et_al_2013}, or even tens to hundreds of pc separations in minor mergers \citep[][]{Callegari_et_al_2011} or in peculiar environments such as in clumpy high-redshift galaxies \citep[][]{Tamburello_et_al_2017}, or in a clumpy gaseous nuclear disc forming after the merger \citep[][]{Roskar_et_al_2015}. Starting from a fully cosmological hydrodynamical simulation \citet{Khan_et_al_2016} succeeded in simulating the decay of a massive BH pair to millipc separations and subsequently to the final merger of BHs. They  extracted a merger between two massive galaxies at $z \sim 3.4$ from a cosmological zoom-in run and re-sampled it at higher mass and force resolution, completing the last evolutionary stage with a collisional $N$-body code, $\phi$\textsc{gpu} \citep{Berczik_et_al_2011}, including post-Newtonian corrections. In the final stage, the ISM was not modelled as most of the gas in the nuclear region had already been consumed by star formation. This led to the first direct determination of the merger time-scale of two BHs in merging galaxies. In this case the merger time-scale was surprisingly short, only 10~Myr after the two galaxy cores coalesced, which was attributed to the very high central baryonic density of the host galaxies due to the fact that they were selected at $z > 3$, aided by the marked triaxiality of the potential \citep[][]{Khan_et_al_2016,Mayer_2017}. As these were simulations of massive  galaxies that would later turn into the central giant elliptical of a rich galaxy group \citep[][]{Feldmann_Mayer_2015}, the BHs also had large masses, $\sim$$10^8$~M$_{\odot}$. As a result, GWs emitted during the inspiral  phase have a very low frequency and would fall marginally inside the LISA frequency window \citep[][]{Mayer_2017}.

In order to ascertain the merger time-scales of BHs whose GW-driven evolution would be well within the LISA band, one needs to consider the dynamical evolution of BHs with lower masses, $<10^7$~M$_{\odot}$. Such BHs reside in disc-dominated galaxies at the present epoch, such as that in our own Milky Way \citep[][]{Greene_Ho_2007,Kormendy_Ho_2013,Greene_et_al_2016}. Presumably this was the case also at higher redshift as the local correlations between the various metrics of galaxy mass and mass of the central BH seem to hold (or mildly evolve) even at higher redshift \citep[][]{Merloni_et_al_2010}.

The merging process of such BHs in disc dominated host galaxies is indeed the focus of this paper. As in \citet{Khan_et_al_2016} we will employ a multi-scale, multi-stage simulation technique to follow the evolution of the BH binary formed after the galaxy merger until it enters the stage of linear hardening in the hard-binary regime. Subsequent evolution and merger times are estimated using constant hardening rates obtained in last phase of the binary evolution in our simulations together with energy loss by GW emission. To limit the computational burden and start with model galaxies with well resolved nuclear mass distribution (at scales less than 100~pc) we employ a subset of the mergers presented in \citeauthor{Capelo_et_al_2015} (\citeyear{Capelo_et_al_2015}; hereafter CAP15; see also \citealt{Capelo_Dotti_2017}) instead of adopting cosmological simulations. The nuclear density profiles in the merger remnants were verified to be very similar to those of disc dominated galaxies formed self-consistently in the Eris suite of cosmological simulations at similar redshifts ($z \sim 2$--3), which were run with nearly identical setup of the smoothed particle hydrodynamic (SPH) code \textsc{gasoline} \citep[][]{Wadsley_et_al_2004} employed in this paper \citep[see, e.g.][]{Bonoli_et_al_2016,Sokolowska_et_al_2017}. Furthermore, the chosen model galaxies have moderate masses of the gas disc and moderate star formation rates, hence they do not develop a clumpy, turbulent ISM such as the massive star forming galaxies at high redshift, which avoids potential dynamical perturbations that might lead to the stalling of the BH pair at large separations, before a bound binary can form \citep[][]{Tamburello_et_al_2017}. BH growth by accretion and their energetic feedback on the surrounding ISM are taken into account until the system becomes gas-poor and the final evolution is computed with the direct $N$-body code.

The paper is organized as follows. In Section~\ref{sec:Numerical_setup}, we describe the numerical setup, including the hydrodynamic simulations of the large-scale mergers which yielded the initial conditions for the direct $N$-body simulations of this work. In Section~\ref{sec:N-body_simulations}, we characterise in detail the structure of the merger remnants (density, geometry, and angular momentum), whereas in Section~\ref{sec:SMBH_binary_formation_and_evolution} we describe the formation and evolution of the BH binary, down to the coalescence of the two BHs. We conclude in Section~\ref{sec:Conclusions}.

\begin{table}[!t]
\vspace{-0.5pt}
\caption{Galaxy merger runs -- initial conditions} 
\begin{tabular}{c c c c c c c c}
\hline
Run & $\theta_1$ & $\theta_2$ & $t^{\prime}_{\rm sel}$ & BH$_1$ & BH$_2$ & $R_{\rm BH-init}$ & $N$ \\
\hline
A  (02) & 0		& 0		& 0.99 & 1.36  & 2.91 & 175.8	& 1.76 \\
B  (03) & $\pi/4$	& 0		& 1.06 & 1.00  & 5.05 & 22.6	& 1.67 \\
C  (04) & $\pi$		& 0		& 1.57 & 1.48  & 3.93 & 46.9	& 1.83 \\
D  (05) & 0		& $\pi$	& 1.22 & 1.18  & 4.59 & 85.1	& 1.77 \\
\hline
\end{tabular}
\vspace{15pt}
\tablecomments{Column~1: Merger run (with the corresponding run number in CAP15 in parenthesis). Column~2: Initial angle between the primary galaxy angular momentum and the global angular momentum in CAP15's simulations. Column~3: Same as Column~3 but for the secondary galaxy. Column~4: Time (in Gyr) of CAP15's simulations at which we chose the ICs for the direct $N$-body runs. Column~5: Mass (in $10^7$~M$_{\odot}$) of the more massive BH at $t^{\prime}_{\rm sel}$. Column~6: Mass (in $10^6$~M$_{\odot}$) of the less massive BH at $t^{\prime}_{\rm sel}$. Column~7: Separation (in pc) between the two BHs at $t^{\prime}_{\rm sel}$. Column~8: Total number of particles (in millions) for the direct $N$-body runs.}
\vspace{10pt}
\label{Table1}
\end{table}


\section{Numerical Setup}\label{sec:Numerical_setup}

\begin{figure*}[!t]
\vspace{2.5pt}
\centerline{
\resizebox{0.95\hsize}{!}{\includegraphics[angle=270]{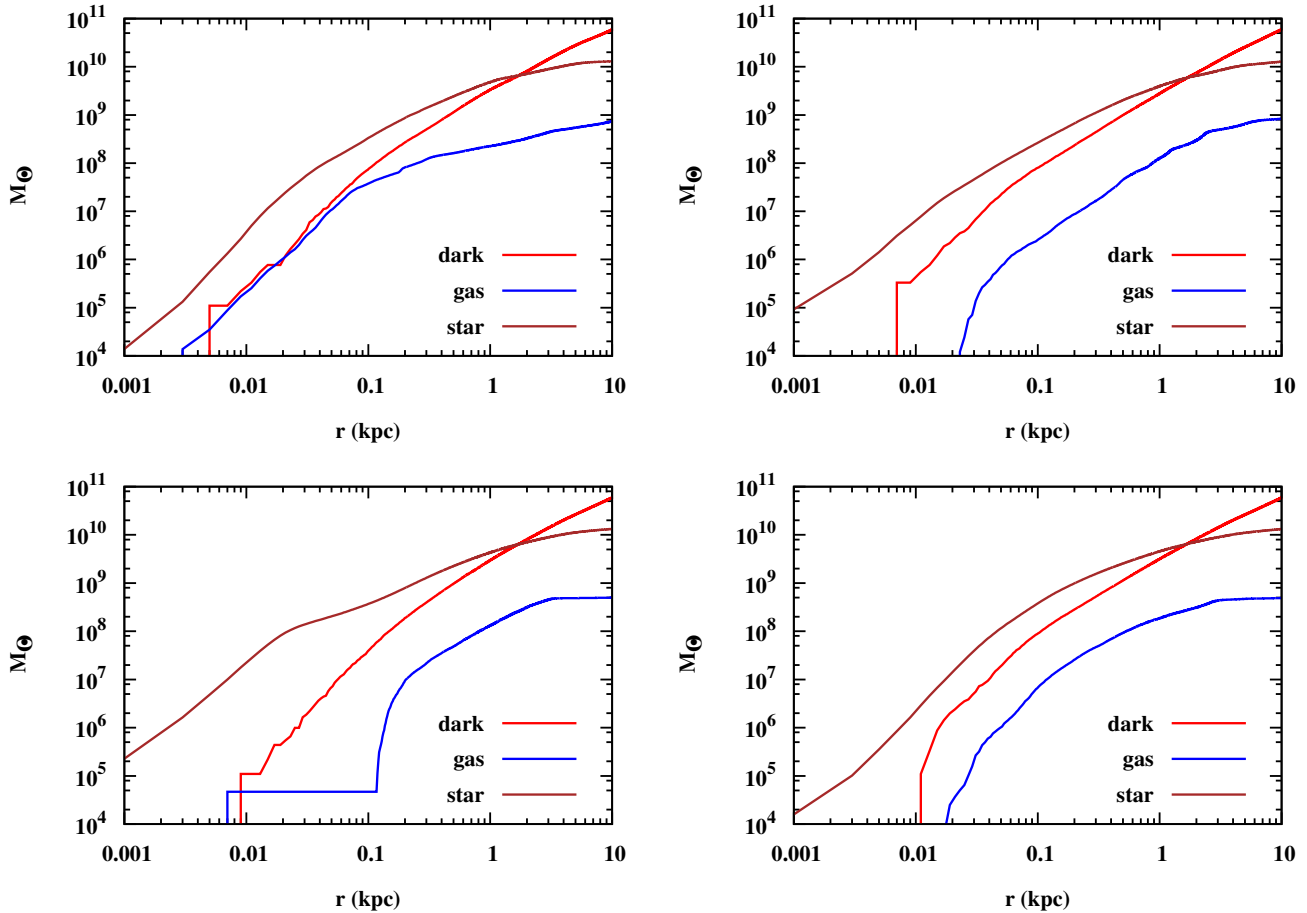}}
}
\vspace{15pt}
\caption{
Cumulative mass profiles for various types of matter at the time of the ICs selection ($t^{\prime} = t^{\prime}_{\rm sel}$; see Table~\ref{Table1}) for all our simulations: Run~A (top-left panel), B (top-right), C (bottom-left), and D (bottom-right).}
\vspace{15pt}
\label{fig:mass-proA}
\end{figure*}

The initial conditions (ICs) for the suite of numerical simulations presented in this study were obtained from the late stages of the galaxy merger simulations of CAP15. In those simulations, late-type galaxies were put at an initial distance equal to the sum of their virial radii and set on parabolic orbits \citep[][]{Benson_2005}, with the distance of the first pericentric passage equal to 20 per cent of the virial radius of the primary galaxy \citep[][]{Khochfar_Burkert_2006}. The angle between the initial individual galactic angular momentum vector of each galaxy and the global angular momentum vector was then varied in order to have coplanar, prograde--prograde, retrograde--prograde, and prograde--retrograde, and inclined encounters (see Columns~3 and 4 of Table~\ref{Table1}).

Each galaxy was composed of a dark matter halo, a baryonic disc (made of stars and gas) and bulge (made of stars), and a central BH. The structural parameters of the simulated galaxies were typical of high-redshift ($z \sim 3$) galaxies \citep[see also discussion in][]{Capelo_et_al_2017}. For the detailed description of all the profiles and parameters, we refer to CAP15.

The suite presented in CAP15 and \citet{Capelo_Dotti_2017} was a follow-up of a similar suite of mergers \citep[][]{Callegari_et_al_2009,Callegari_et_al_2011,VanWassenhove_et_al_2014} which was also constructed to study the pairing time-scales of BHs in unequal-mass galaxy mergers. In all those simulations, the gravitational softening of all the particles was of the order of 10--30~pc, but see \citet{Pfister_et_al_2017} for a recent higher-resolution SPH study of some of the same mergers.

Out of the complete set of CAP15, we chose runs with the same initial mass ratio (1:2). We selected all the particles within a sphere of radius 3~kpc around the BHs' centre of mass, when the separation between the BHs was a few times greater than 20~pc, of the order of the spatial resolution of CAP15's simulations. Table~\ref{Table1} gives the parameters of our ICs for the selected runs. Figure~\ref{fig:mass-proA} shows the cumulative mass distribution of dark matter, gas, and stars for all our runs at the time of our selection. We note that the stellar mass dominates both over the gaseous and dark matter components in the centre ($<$100~pc) by more than an order of magnitude in all cases. Since the stellar mass dominates over the gas mass for all the models (in contrast to an initial gas fraction of 30 per cent at the beginning of the corresponding simulations in CAP15), we treated the residual gas particles as stellar particles. However, the total number of stellar particles $N_{\star}$ for all the models selected in this way resulted to be roughly $3 \times 10^6$, which is a large number for direct $N$-body simulations, especially when one wants to perform a set of them as we did in this study. Therefore, we reduced $N_{\star}$ by a factor of two by deleting each second star in our sample and adding its mass to the surviving one, as it was shown that a change of a factor of two in the number of particles does not affect the results \citep[see, e.g.][]{Khan_et_al_2011,pre11}. This way we got $N_{\star} \simeq 1.6$--$1.7 \times 10^6$ which, by adding $\sim$ $10^5$ dark matter particles, resulted in a total $N \simeq 1.7$--$1.8 \times 10^6$.

The gravitational softenings employed in the simulations of CAP15 were 10, 20, and 30~pc for stars, gas, and dark matter, respectively. We increased the dark matter softening to 50~pc for the direct $N$-body simulations, to avoid occasional strong interactions between dark matter particles and BHs, which have an average mass contrast of roughly 114 and 38 for the primary and secondary BH, respectively. For the stellar particles, we reduced the softening to 0.1~pc to follow the three-body hardening phase of hard BH binaries consistently. The average mass contrast for stellar particles and BHs is $3.8 \times 10^3$ and $1.2 \times 10^3$ for the primary and secondary BH, respectively. The initial masses of the primary and secondary BH at the start of the simulations of CAP15 (for the subset of simulations presented here) were $3.53 \times 10^6$ and $1.77 \times 10^6$~M$_{\odot}$, respectively. The BH masses increased depending on the gas accretion history caused by various configurations of galaxy mergers in the previous phase of hydrodynamic simulations such that, at the time of our selection, the BH masses increased by factors of 2--4 (see Table~\ref{Table1}).\\\\


\section{Direct $N$-body simulations}\label{sec:N-body_simulations}

\begin{figure*}[!t]
\vspace{-13pt}
\centering
\vspace{3pt}
\begin{overpic}[width=1.02\columnwidth,angle=0]{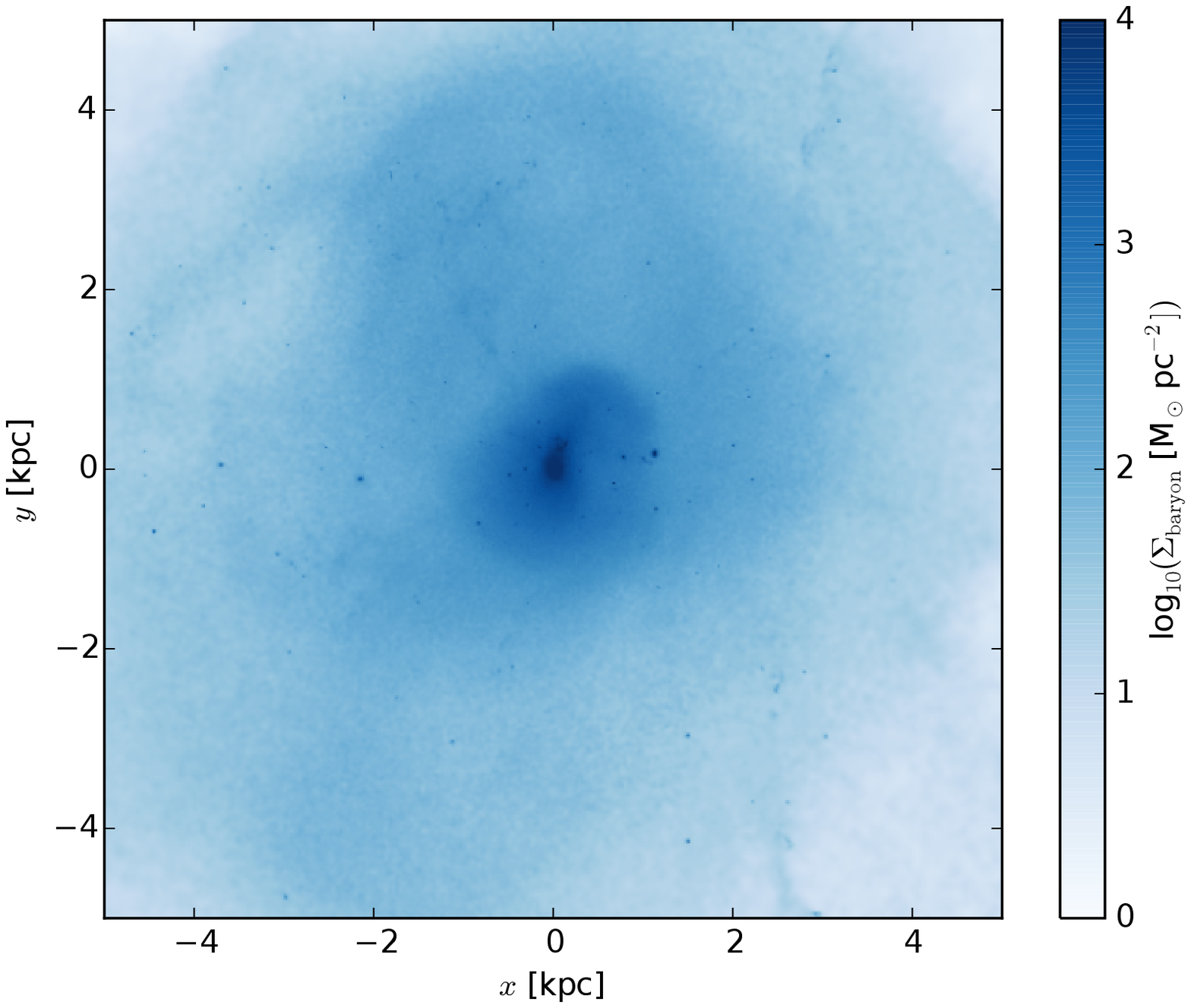}
\put (16,63) {\textcolor{black}{A}}
\end{overpic}
\begin{overpic}[width=1.02\columnwidth,angle=0]{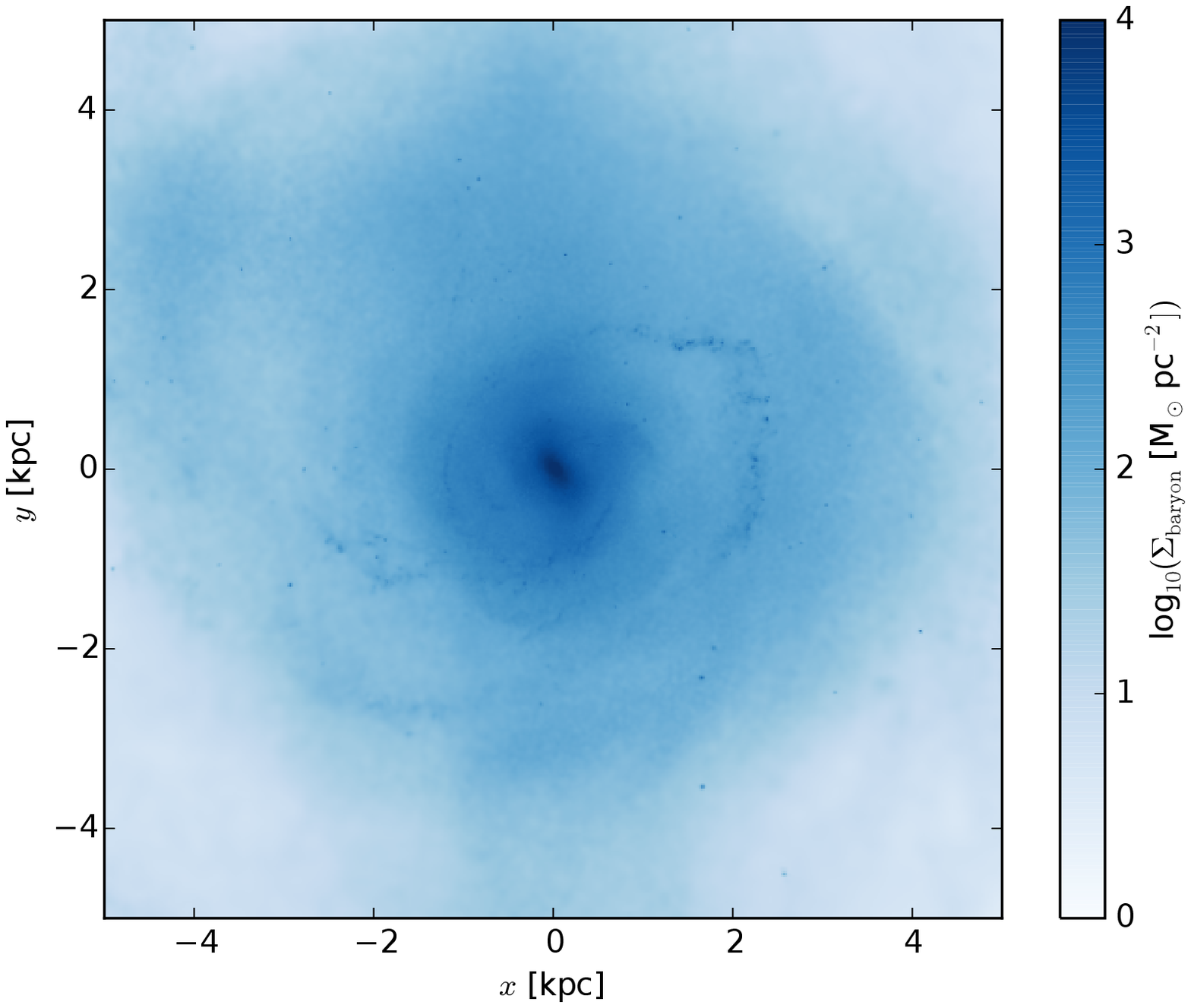}
\put (16,63) {\textcolor{black}{B}}
\end{overpic}
\vskip 0.0mm
\begin{overpic}[width=1.02\columnwidth,angle=0]{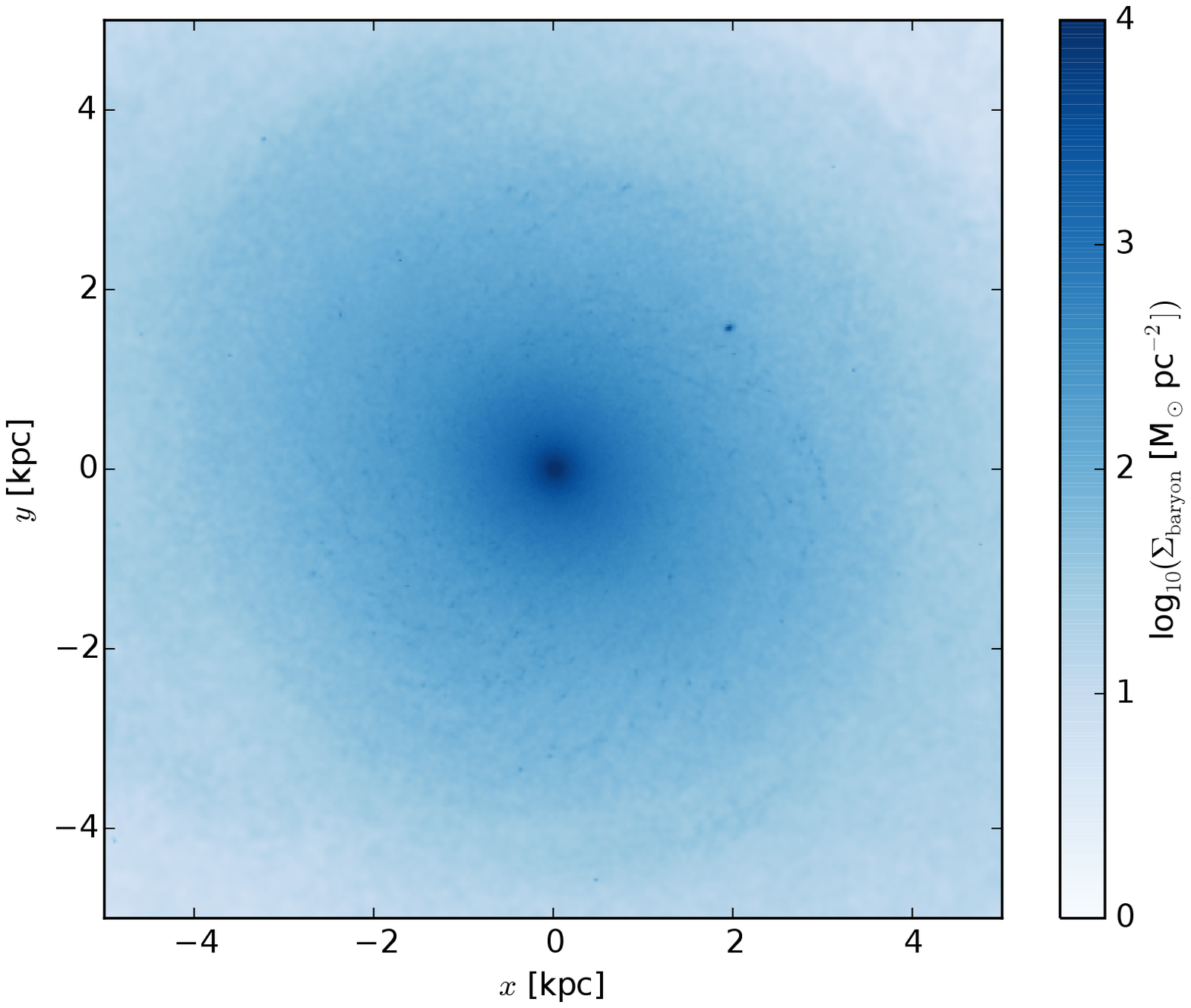}
\put (16,63) {\textcolor{black}{C}}
\end{overpic}
\begin{overpic}[width=1.02\columnwidth,angle=0]{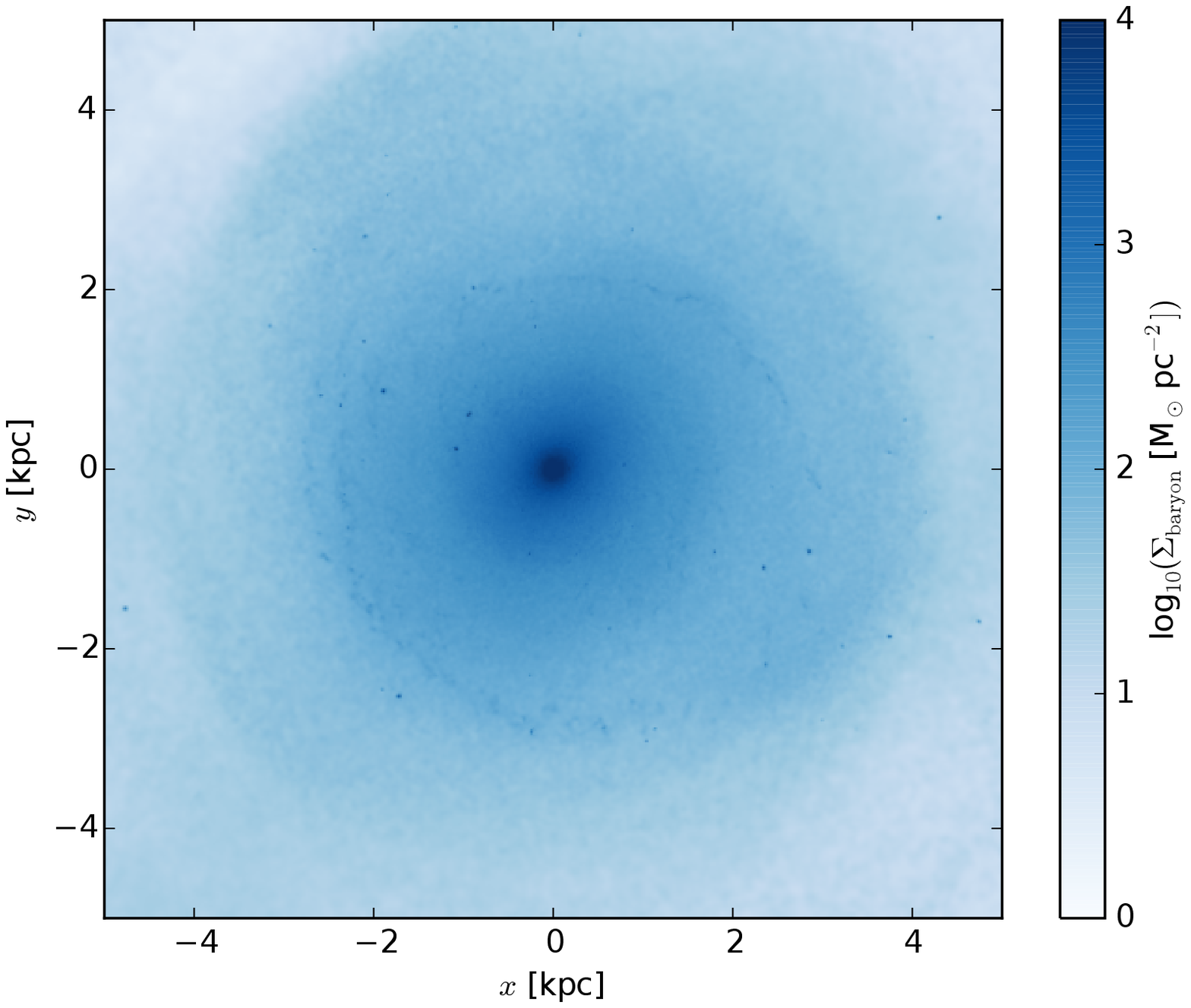}
\put (16,63) {\textcolor{black}{D}}
\end{overpic}
\vspace{15pt}
\caption{Baryonic density snapshots (viewed face-on) of the central region at the time of the ICs selection ($t^{\prime} = t^{\prime}_{\rm sel}$; see Table~\ref{Table1}) for all our simulations: Run~A (top-left panel), B (top-right), C (bottom-left), and D (bottom-right).}
\vspace{15pt}
\label{fig:faceon2}
\end{figure*}

The extracted central region of the galaxy mergers, as described in the previous section, is further evolved using the direct $N$-body code $\phi$\textsc{gpu}. At the beginning of our direct $N$-body simulations ($t^{\prime} = t^{\prime}_{\rm sel}$; $t \equiv t^{\prime} - t^{\prime}_{\rm sel} = 0$), the galaxies are already merged (see Figure~\ref{fig:faceon2}) and the BH separations are only a factor of a few influence radii $r_{\rm infl} \sim 10$--30~pc), computed by finding the distance from the centre of mass of the two BHs at which the enclosed stellar mass is twice the combined mass of the BHs. Hence, in all our simulations, we form BH-binary systems soon after the start of our runs. Here we present some useful parameters of our product galaxies.

\subsection{Density Profiles}\label{sec:Density_profiles}

\begin{figure*}[!t]
\vspace{3pt}
\centerline{
\resizebox{0.95\hsize}{!}{\includegraphics[angle=270]{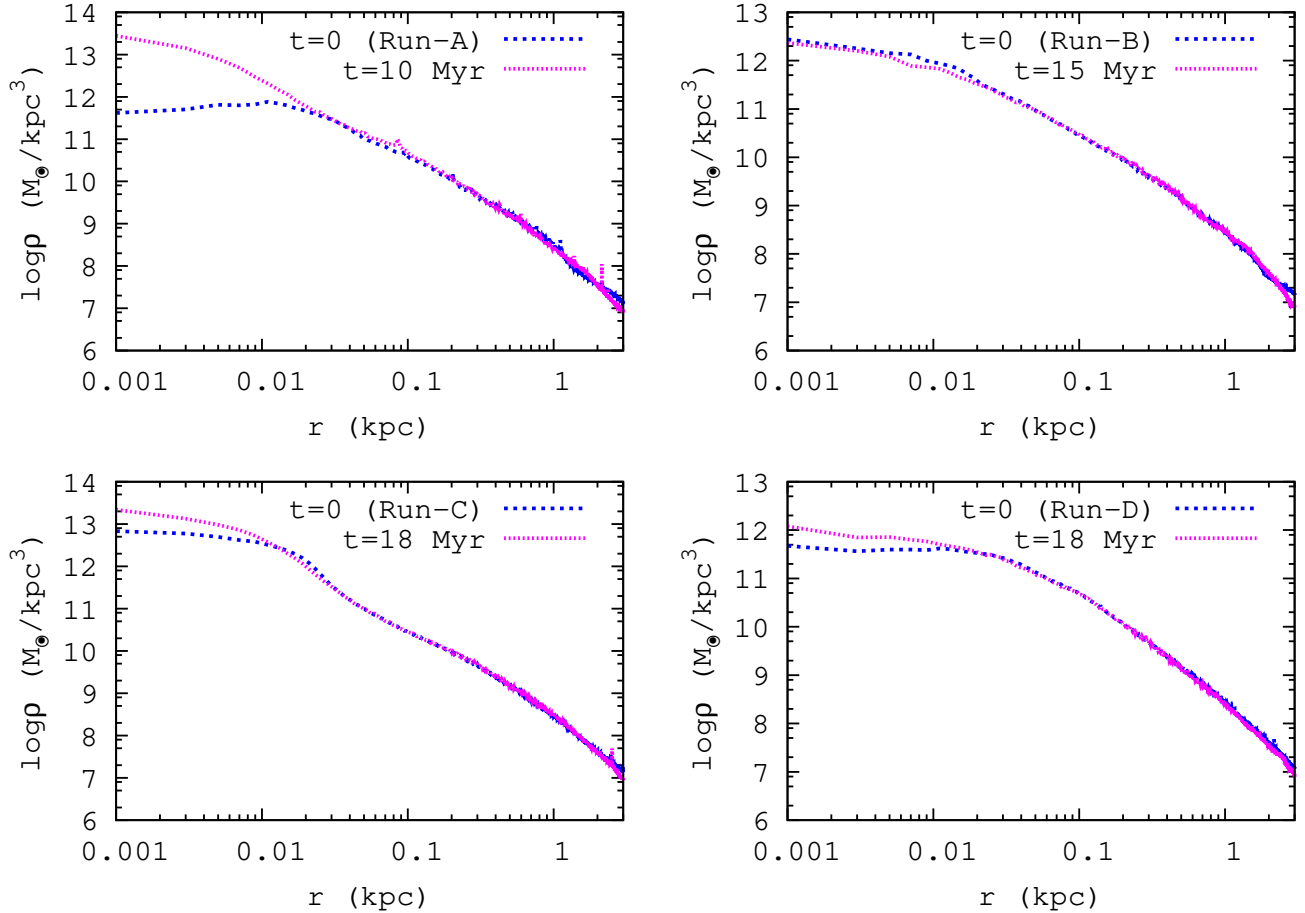}}
}
\vspace{15pt}
\caption{
Stellar volume density profiles for our merger simulations at the beginning of the direct $N$-body simulations ($t^{\prime} = t^{\prime}_{\rm sel}$; see Table~\ref{Table1}) and at a later time $t = t_{\rm hard} = 10$--18~Myr (depending on the run; see text), for all our simulations: Run~A (top-left panel), B (top-right), C (bottom-left), and D (bottom-right). 
}
\vspace{15pt}
\label{fig:multidens}
\end{figure*}

We calculate the volume density distribution for the stars centred on the BH pair's centre of mass at the start of our simulations (when the distribution is identical to that of the hydrodynamic simulations) and compare it to that at a later time $t = 10$--18~Myr (depending on the run) in the direct $N$-body runs. Times are chosen during an interval when a hard Keplerian binary evolves in the three-body scattering phase of BH binary evolution. More specifically, we check when the BH separation reaches the hard-binary separation $a_{\rm h}$, defined as \citep[][]{Merritt_2013} $M_{\rm BH_2} r_{\rm h}/[4(M_{\rm BH_1}+M_{\rm BH_2})]$, where $r_{\rm h}$ is the influence radius of the larger BH, for which we take as proxy $r_{\rm infl}$. We choose these later times (which we call $t = t_{\rm hard}$\footnote{In our notation $t = t_{\rm hard}$ is not the time of formation of a hard BH binary, rather it is an arbitrary time of selection of snapshots for analysis in hard binary regime.}) for the analysis of the density profiles because, during the interval from binary formation to hard-binary formation, the central stellar density drops drastically due to core scouring by the massive binary \citep{Merritt_2006,Khan_et_al_2012a,Rantala_et_al_2018}. Figure~\ref{fig:multidens} shows the result for all four merger runs. The stellar density profiles of the direct $N$-body and hydrodynamic simulations are very similar except at the very centre. The two profiles differ as expected inside $\sim$20~pc, the gas softening used in CAP15. We witness a mild increase in density towards the centre in the direct $N$-body simulations except in Run~A, which has more than an order of magnitude increase. Overall, Runs~A and C have comparable central densities, significantly higher than those of Runs~B and D. The central density and the stellar distribution geometry play a critical role in affecting the hardening rates and hence driving BH coalescence via GW emission \citep[][]{Khan_et_al_2012a}.

 
\subsection{Merger Remnant Geometry}\label{sec:Merger_remnant_geometry}

The shape of the merger remnant is a key factor to avoid the so-called final-parsec problem \citep{Merritt_Poon_2004}. We calculated the triaxiality parameter $T$, defined as 

\begin{equation}\label{eq:triaxiality}
T = \dfrac{(b-c)}{(a-c)},
\end{equation}

\noindent where $a$, $b$, and $c$ are the major, intermediate, and minor axes calculated for a uniform ellipsoid from the inertia tensor. The results for the triaxiality parameter are shown for the stellar and dark matter distributions in Figure~\ref{fig:triax}. It appears that the stellar distribution in the central kpc has a strongly triaxial shape for all merger runs except for Run~C, which has a mild triaxiality. The dark matter distribution appears to exhibit an even stronger triaxiality for all the runs. Triaxial stellar and dark matter distributions in the central kpc strongly suggest that the BH binary evolution in such merger remnants should happen independently of $N$, without experiencing the final-parsec problem \citep{Khan_et_al_2011,Rantala_et_al_2017}.


\subsection{Merger Remnant Angular Momentum}\label{sec:Merger_remnant_angular_momentum}

BH binary dynamics can depend strongly on the alignment/counter-alignment of the BH binary and galaxy angular momenta \citep[][]{Sesana_et_al_2011,HolleyBockelmann_Khan_2015,Mirza_et_al_2017}. In Figure~\ref{fig:multimomgal}, we plot the normalised angular momentum components of the stellar component of the post merger remnant, calculated in spherical shells of radius 20~pc around the centre of mass of the BH binary. We notice that the angular momentum of the merger remnant is dominated by the initial angular momentum of the primary galaxy. For Runs~A and D, the primary galaxy has its angular momentum in the $z$ direction ($\theta = 0$) and so do the stellar mass distributions in the merger remnants. For Run~C, the angular momenta of the primary galaxy and merger remnant are in the $-z$ direction ($\theta = -\pi$~radians), whereas for Run~B, where the primary galaxy is inclined at an angle $\theta = \pi/4$~radians, the merger remnant has mixed values of angular momentum components, albeit with a dominant component in the $x$ direction.\\\\


\section{Supermassive Black Hole Binary Formation and Evolution}\label{sec:SMBH_binary_formation_and_evolution}

In this section, we present the plots for various BH binary parameters. Labels are as in Table~\ref{Table1}.


\subsection{BH Separation Evolution}\label{sec:BH_separation_evolution}

\begin{figure*}[!t]
\vspace{0pt}
\centering
\vspace{3pt}
\begin{overpic}[width=1.32\columnwidth,angle=270,trim={0 0 0.5cm 0},clip]{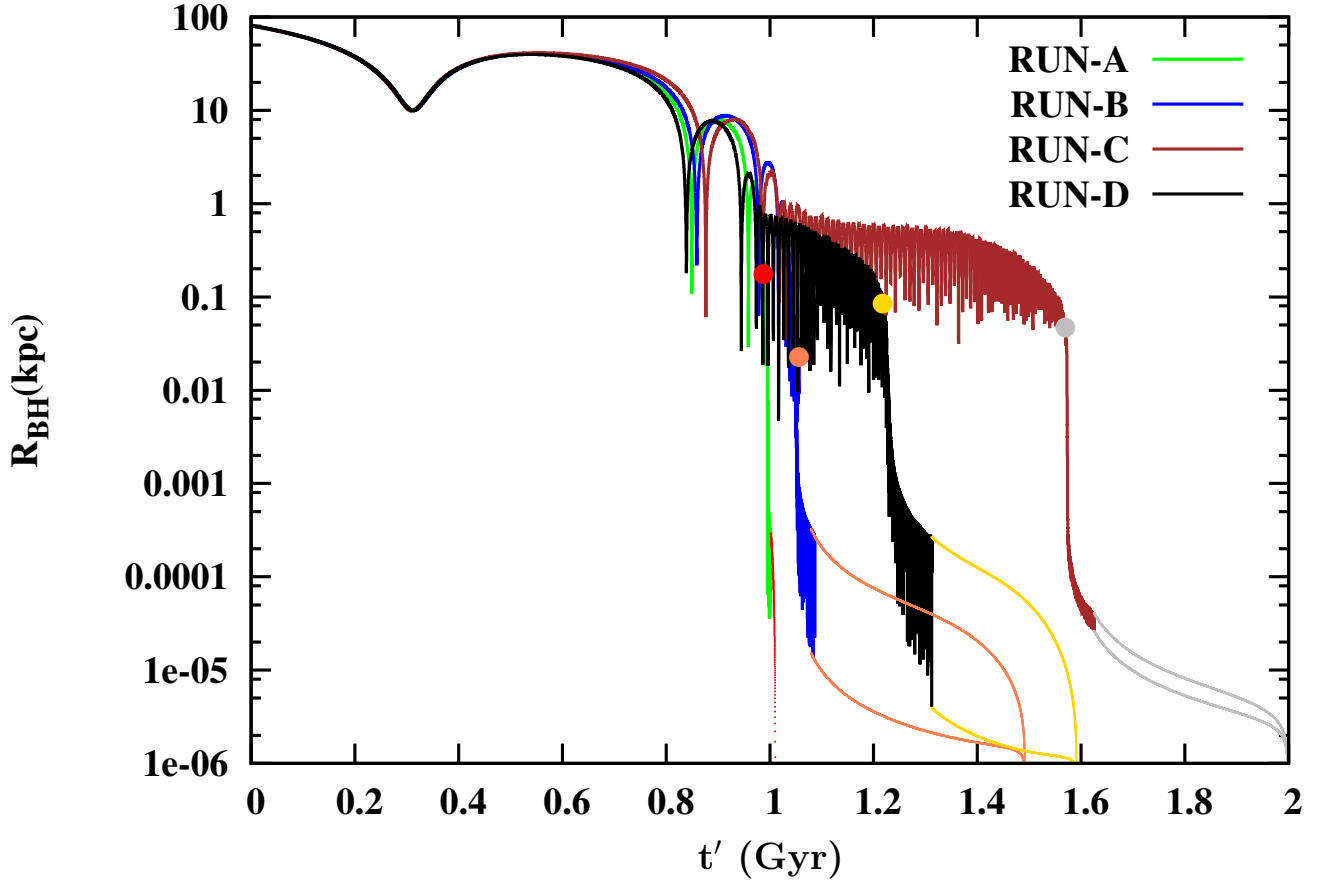}
\put (53,-2.5) {\textcolor{black}{\Large \bf t$^{\prime}$ (Gyr)}}
\end{overpic}
\vspace{15pt}
\caption{Relative separation between the BHs from the beginning of the hydrodynamic simulations of CAP15, then through the three-body scattering phase simulated with the direct $N$-body runs, till the estimated merger of BHs. The galactic remnant in the hydrodynamic simulations forms at different times, depending on the encounter, but always in the range 1--1.1~Gyr. The beginning of the direct $N$-body simulations is highlighted by the filled circles. The estimated evolution is computed by choosing binary parameters when we stop direct N-body simulations.
}
\vspace{15pt}
\label{fig:seplucio1}
\end{figure*}

The BH separation evolution during the course of each galaxy merger and subsequent BH binary hardening phase is shown in Figure~\ref{fig:seplucio1}. The transition from the hydrodynamic simulations of CAP15 to the direct $N$-body simulations of this study is shown by filled circles for all the runs. We note that the BH separation shrinks by almost two orders of magnitude in about ten Myr after the transition. This rapid phase of BH separation shrinking is governed jointly by dynamical friction and three-body encounters of stars with the BH binary (as the BHs form a Keplerian binary). Later on, as the BH binary erodes the surrounding stellar cusp, dynamical friction becomes inefficient and the BH separation shrinks at a slower and almost constant rate in the three-body hardening regime.   

\begin{figure}[!t]
\vspace{2pt}
\centerline{
\resizebox{0.95\hsize}{!}{\includegraphics[angle=270]{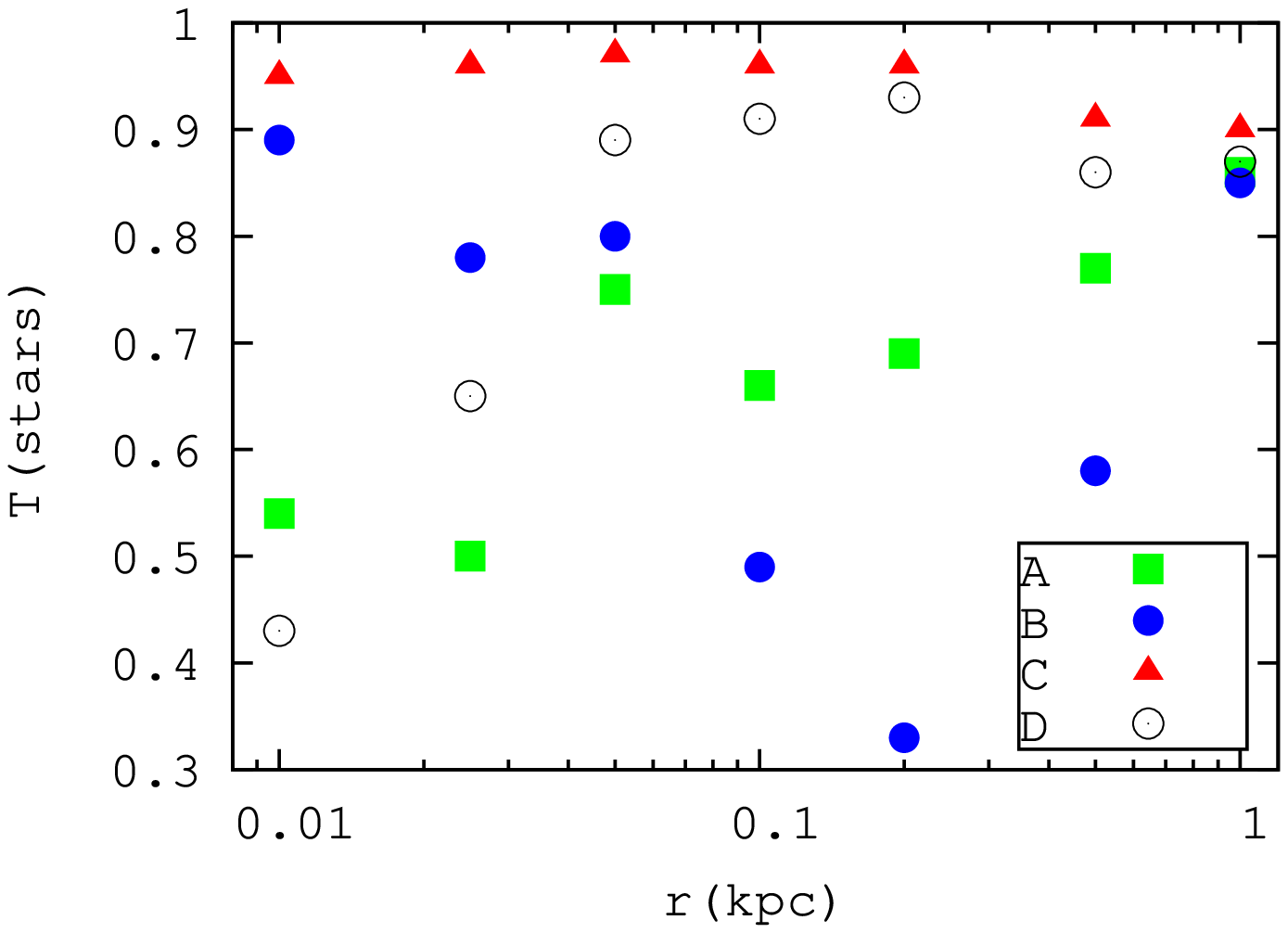}}
}
\centerline{
\resizebox{0.95\hsize}{!}{\includegraphics[angle=270]{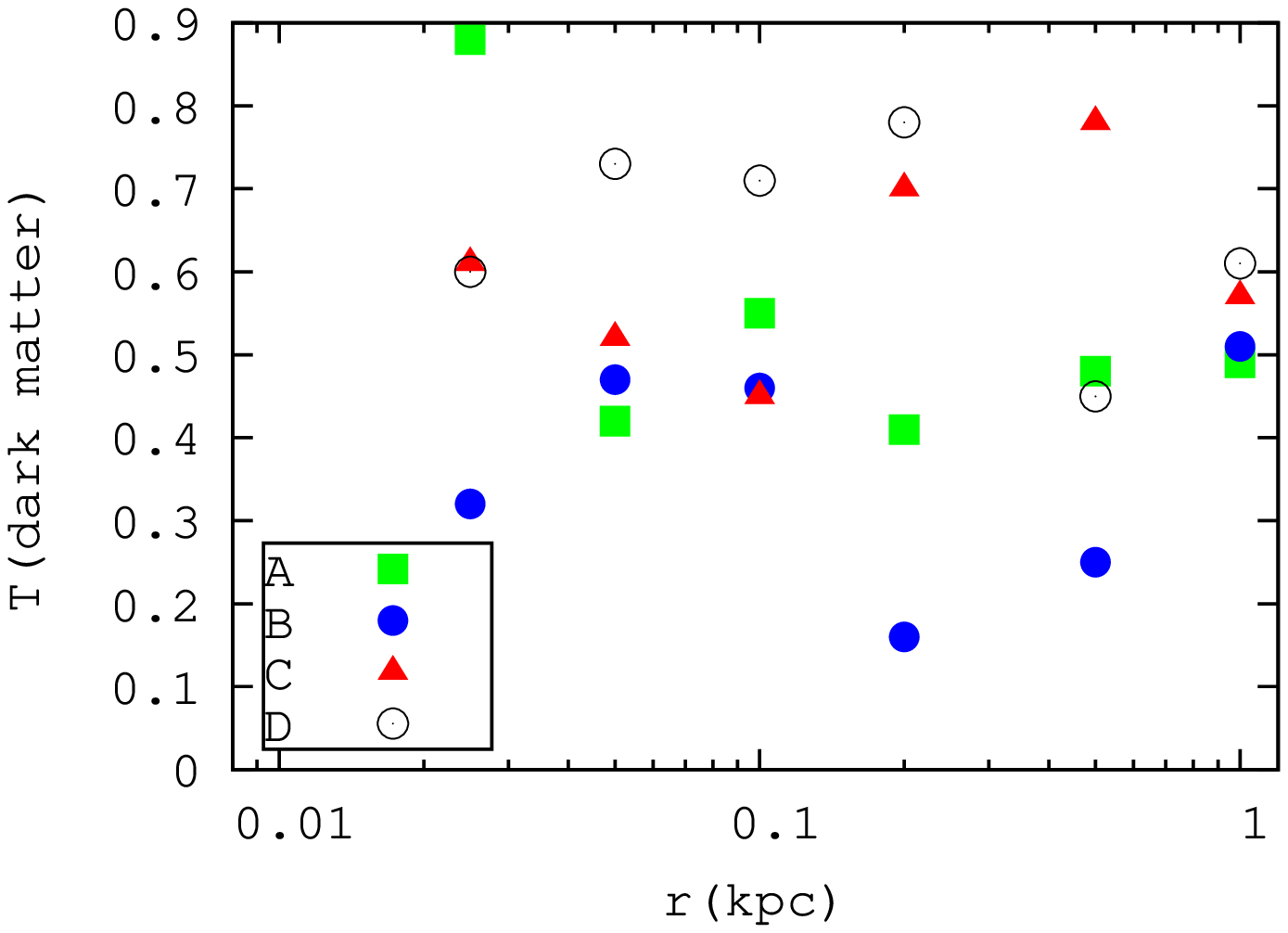}}
}
\vspace{15pt}
\caption{
Radial triaxiality profiles for the stellar (top panel) and dark matter (bottom) distributions, measured at $t = t_{\rm hard}$.
}
\vspace{15pt}
\label{fig:triax}
\end{figure}
     
\begin{figure*}[!t]
\vspace{3.5pt}
\centerline{
\resizebox{0.9\hsize}{!}{\includegraphics[angle=270]{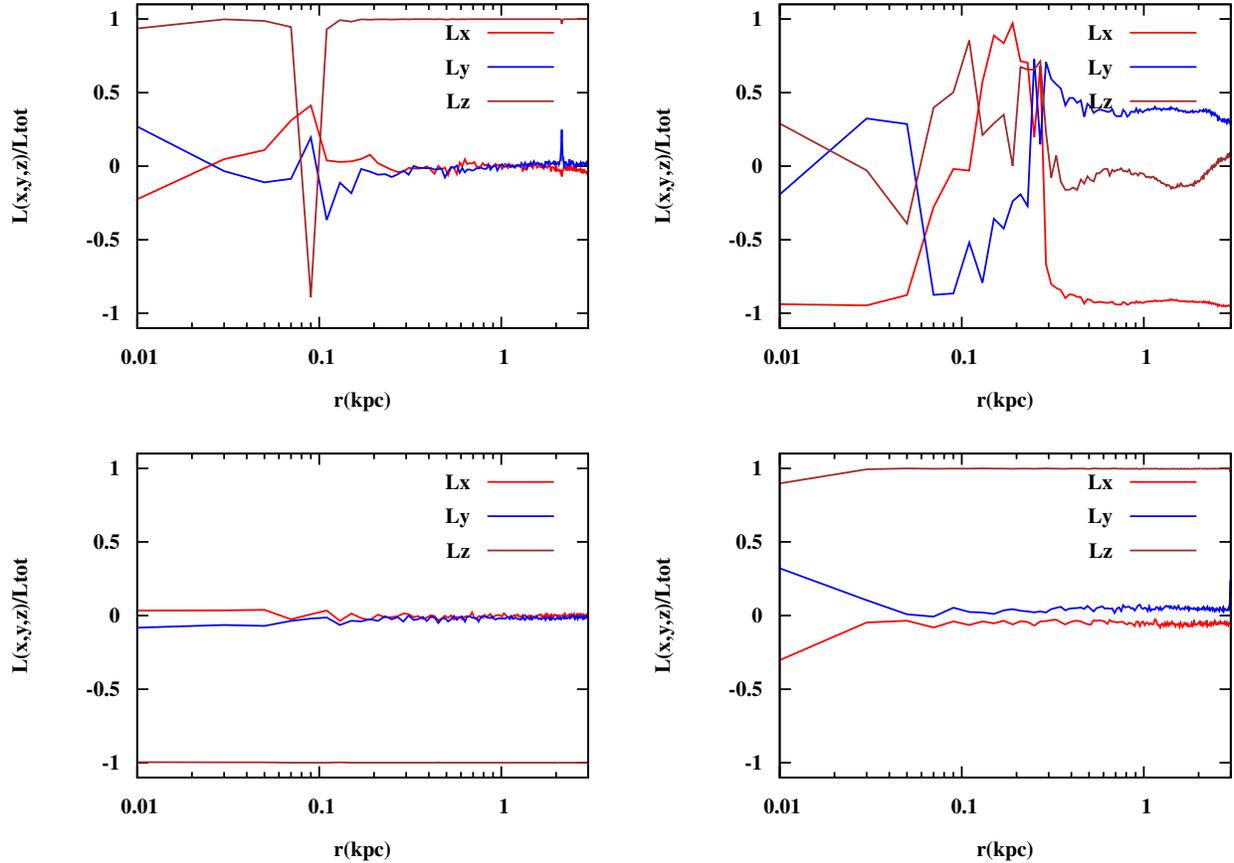}}
}
\vspace{15pt}
\caption{
Radial angular momentum profiles for the merger remnants, measured at $t = t_{\rm hard}$, for Runs~A (top-left panel), B (top-right), C (bottom-left), and D (bottom-right).
}
\vspace{15pt}
\label{fig:multimomgal}
\end{figure*}

\subsection{BH Binary Semi-Major Axis Evolution}\label{sec:BH_binary_semimajor_axis_evolution}

\begin{table}[!t]
\vspace{-0.5pt}
\caption{Galaxy merger runs -- final properties} 
\begin{tabular}{c c c c c c c c c}
\hline
Run & $r_{\rm infl}$ & $a_{\rm h}$ & $\rho_{\rm \star cen}$ & $\rho_{\rm \star infl}$ & $s$ & $e$ & $t_{\rm coal}$ & $t^{\prime}_{\rm coal}$\\
\hline
A & 13 & 0.57 & $27$ & $14$ & 723 & 0.99 & 0.025 & 1.02\\
B & 26 & 2.18 & $3.1$ & $2.4$ & 158 & 0.91 & 0.44 & 1.45\\
C & 19 & 1.00 & $21$ & $11$ & 665 & 0.11 & 0.42 & 1.99\\
D & 27 & 1.89 & $1.2$ & $3$ & 85 & 0.93 & 0.29 & 1.51\\
\hline
\end{tabular}\label{Table2}
\vspace{15pt}
\tablecomments{Column~1: Merger run (see Table~\ref{Table1}). Columns~2 and 3: Influence radius and hard-binary separation (in pc) of the BH binary, calculated at $t = t_{\rm hard}$. Columns 4 and 5: central stellar volume density (in $10^{12}$~M$_{\odot}$~kpc$^{-3}$) and stellar volume density (in $10^{11}$~M$_{\odot}$~kpc$^{-3}$) at the influence radius, respectively, computed at $t = t_{\rm hard}$. Column~5: BH binary hardening rate (in kpc$^{-1}$~Myr$^{-1}$), computed in the late phase of the binary evolution. Column~6: BH binary eccentricity, computed at the end of the direct $N$-body simulation. Column~7: Approximate BH merger time (in Gyr), from the start of the direct $N$-body run. Column~8: Total BH merger time (in Gyr), from the start of the hydrodynamic simulations.}
\vspace{15pt}
\end{table}

\begin{figure}[!t]
\vspace{1.5pt}
\centerline{
\resizebox{0.95\hsize}{!}{\includegraphics[angle=270]{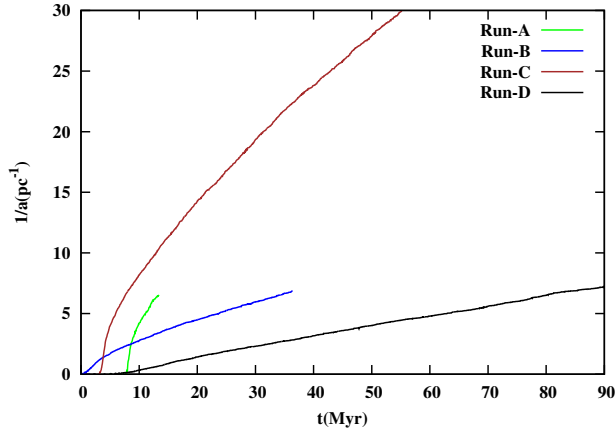}}
}
\vspace{10pt}
\caption{
Inverse semi-major axis of the (Keplerian) BH binaries. The hardening rate $s$ is computed during the late phase of the binary evolution.
}
\vspace{10pt}
\label{fig:bbhsemi}
\end{figure}

The BH binary inverse semi-major axis $1/a$ is plotted for our direct $N$-body runs in Figure~\ref{fig:bbhsemi}. Runs~A and C have a steep time evolution of $1/a$, whereas Runs~B and D have a relatively slow growth rate. We calculated the hardening rate $s = {\rm d}(1/a)/{\rm d}t$ by determining the slope of the inverse semi-major axis growth line fitted by a straight line during the linear phase of evolution. We see from Table~\ref{Table2} that Runs~A and C have hardening rates roughly 5--10 times higher than those for Runs~B and D. As the BH masses are of the same order in all runs, these high hardening rates in Runs~A and C should be caused by higher central densities in the merger remnant for these cases \citep[][]{Khan_et_al_2012a}, in accordance to the relation

\begin{equation}\label{eq:hardening}
s = \frac{GH\rho}{\sigma},
\end{equation}

\noindent where $H \approx 16$ is a dimensionless hardening parameter, and $\rho$ and $\sigma$ are the stellar density and velocity dispersion, usually taken at the influence radius \citep[][]{Sesana_Khan_2015}. Indeed this is evident from the density values in Table~\ref{Table2}, both at the centre and at the influence radius, which are roughly 5--10 times higher in Runs~A and C than in Runs~B and D. The density difference is expected, since the efficiency of merger-induced torques is maximised in coplanar, prograde--prograde mergers, leading to stronger gas and stellar inflows (e.g. \citealt{Cox_et_al_2008}; CAP15). Moreover, the strength of the interaction between the two gas discs is higher in coplanar mergers than in inclined mergers, also leading to increased gas inflows and concurrent star formation \citep[][]{Capelo_Dotti_2017}.


\subsection{BH Binary Eccentricity Axis Evolution}

The simulated BH binaries of Runs~A, B, and D form with high values\footnote{We neglect the initial noisy behaviour, due to the fact that in such phase the system is still not completely Keplerian, due to the presence of bound cusps around the individual BHs, which erode with time as the BH binary hardens.} of eccentricity $e$ and reach even higher values ($e > 0.9$) during the three-body scattering phase (see Figure~\ref{fig:eccbbh}). Run~C, on the other hand, starts with low values of eccentricity ($e < 0.1$) and grows gradually to $e \simeq 0.2$.

We try to explain the behaviour of eccentricity in light of the findings of \citet{Sesana_et_al_2011} and \citet{HolleyBockelmann_Khan_2015}, who noticed that counter-rotating binaries reach very high values of $e$, whereas co-rotation leads to low BH binary eccentricities. To do so, we plot the angular momentum components of the BH binaries in Figure~\ref{fig:multimombh} and compare them with the galaxy angular momentum components plotted in Figure~\ref{fig:multimomgal}.

We note that the BH binary plane undergoes random oscillations in Run~A and it is difficult to infer a particular sense of co- or counter-rotation with respect to the host galaxy. For Run~B, the dominant angular momentum component of the galaxy is in the negative $x$-direction, whereas the BH binary has a dominant component in the positive $x$-direction. Hence, for Run~B, we witness a counter-rotation scenario and a high value of eccentricity, consistent with expectations. For Run~C, we see a clear scenario of co-rotation, with both the dominant components of the galaxy and BH binary angular momentum aligned in the negative $z$-direction, and a low value of eccentricity, again consistent with \citet{Sesana_et_al_2011} and \citet{HolleyBockelmann_Khan_2015}. For Run~D, the BH binary's orbital plane constantly changes (especially during the first 20~Myr), as was the case for Run~A. Therefore, the eccentricity behaviour of BH binaries witnessed in isolated rotating systems seems to work in realistic merger situations. Additionally, we notice that if the BH binary's orbital plane is unstable, as is the case for Runs~A and D, then it can cause high values of eccentricity.


\subsection{Estimated Merger Time of BH Binaries}\label{sec:Estimated_merger_time_of_BH_binaries}

We estimated the merger time of BH binaries in our simulations by extrapolating a constant hardening rate $s$ in the stellar dynamical hardening regime, coupled with \citet{Peters_Mathews_1963}'s leading order equations for energy loss by orbiting masses due to GW emission \citep[e.g.][]{Khan_et_al_2012a,Sesana_Khan_2015}. It was shown in our earlier study \citep{Khan_et_al_2012a} that such estimates match reasonably well with merger times obtained by post Newtonian simulations incorporating terms up to 3.5 order. The estimated evolution is shown in Figure~\ref{fig:seplucio1} and the estimated merger times are listed in Table~\ref{Table2}. We see that the longest phase is the galaxy merger phase, which takes a little more than 1~Gyr, and that the BH merger happens efficiently in a few hundred Myr after the galaxies merge. Run~B is an exception, wherein the BHs coalesce in almost radial orbits just after the formation of a hard BH binary.

\begin{figure*}[!t]
\vspace{15pt}
\centerline{
\resizebox{0.95\hsize}{!}{\includegraphics[angle=0]{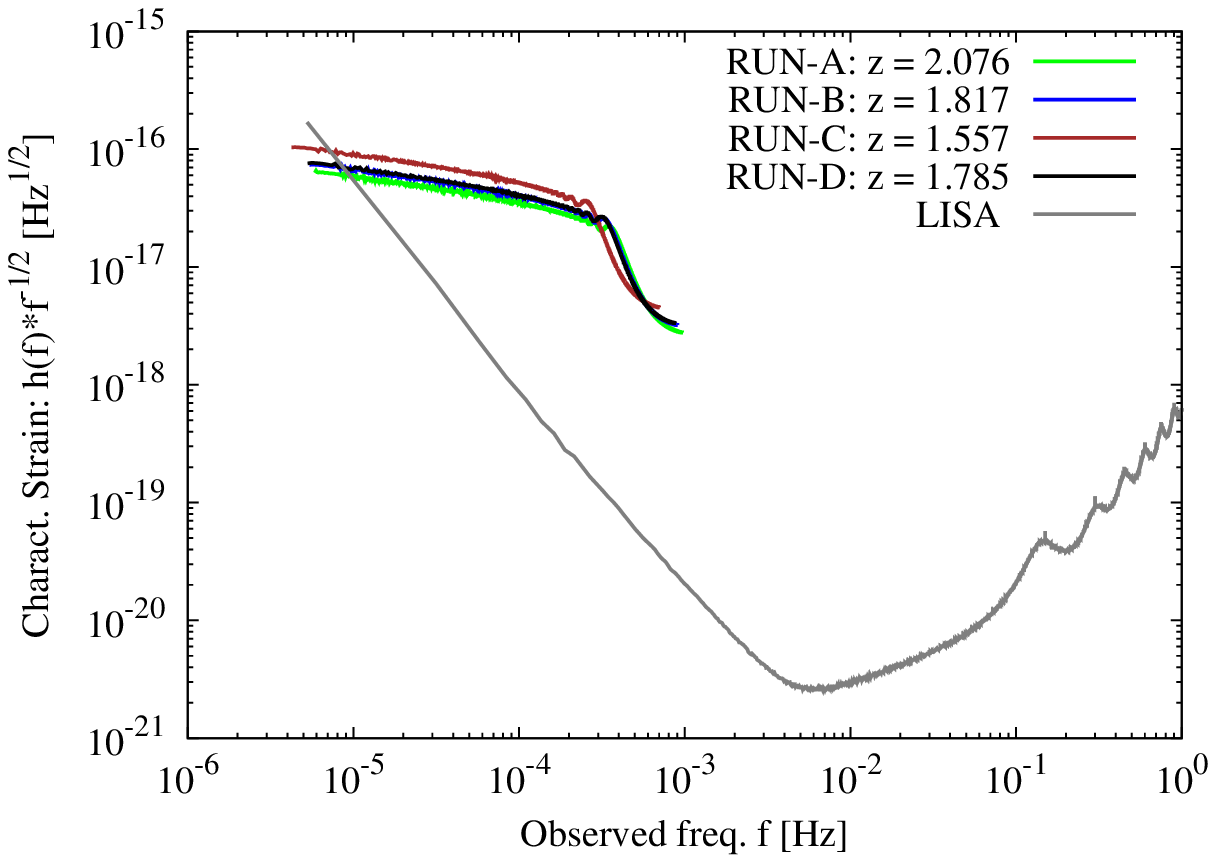}}
}
\vspace{15pt}
\caption{Characteristic strain for all our BH mergers at corresponding $z$ and LISA sensitivity curve.
}
\vspace{10pt}
\label{gwstrain}
\end{figure*}

We also calculate the characteristic strain for all BH merger cases, using estimated parameters at the redshift corresponding to our calculated merger time (Column~8 of Table~\ref{Table2}), assuming that $t^{\prime} = 0$ corresponds to $z = 3$. 
The strain signal is calculated using two body Hermite 4$^{\rm th}$ order Post Newtonian code \citep[][]{SBSK2017,Berczik_et_al_2011,BSW2013} which calculates the orbital evolution of the SMBH GW merger up to the separation of the last few Schwarzschild radius.  LISA sensitivity curve is plotted in accordance with  \citet[][]{LISA2017,MCB2015} and a very helpful online GW plotting page {\tt http://gwplotter.com/}. The final results are plotted in figure~\ref{gwstrain} containing the last 
few months of physical time of the BH binary orbital evolution before the final merger. We see that mergers of BH for all our cases fall well within the observable window of LISA \citep{Amaro-Seoane_et_al_2013,GOAT_2016,Barack_et_al_2018}.





\section{Conclusions}\label{sec:Conclusions}

\begin{figure}[!t]
\vspace{10pt}
\centerline{
\resizebox{0.95\hsize}{!}{\includegraphics[angle=270]{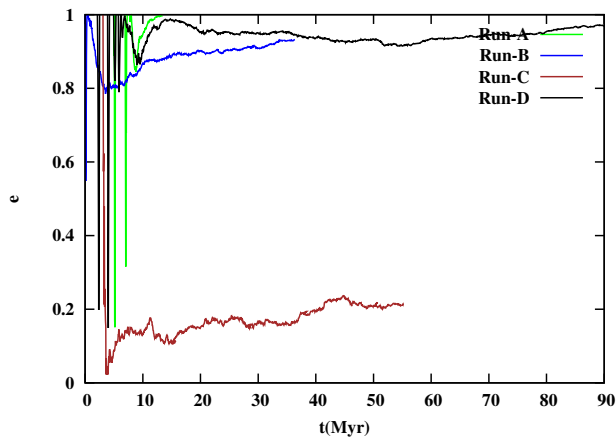}}
}
\vspace{10pt}
\caption{
Eccentricity of the (Keplerian) BH binaries. 
}
\vspace{10pt}
\label{fig:eccbbh}
\end{figure}

\begin{figure*}[!t]
\vspace{10pt}
\centerline{
\resizebox{0.95\hsize}{!}{\includegraphics[angle=270]{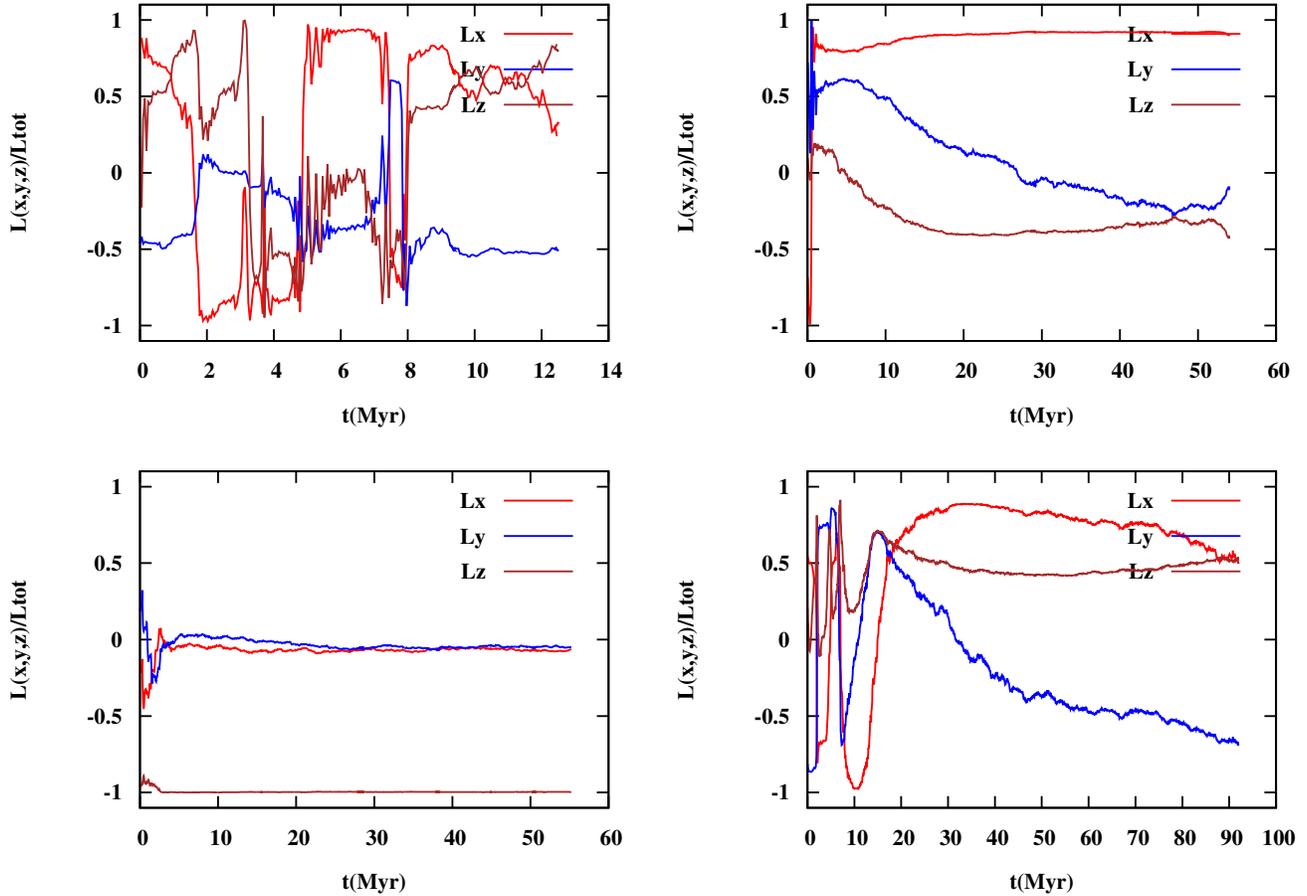}}
}
\vspace{15pt}
\caption{Angular momentum evolution for the BH binaries, calculated for a Keplerian binary, for Runs~A (top-left panel), B (top-right), C (bottom-left), and D (bottom-right).
}
\vspace{15pt}
\label{fig:multimombh}
\end{figure*}

We performed a suite of direct $N$-body simulations of the central regions of late-type galaxy merger remnants, focussing on the fate of the two central BHs. The initial conditions of these simulations were taken from the outputs of four high-resolution SPH simulations (described in CAP15), at a time when a merger remnant has already formed (Figure~\ref{fig:faceon2}) and when gas is extremely sub-dominant (Figure~\ref{fig:mass-proA}). The direct $N$-body simulations employed in this study cover the formation of a BH binary, initially caused by dynamical friction, following up its evolution in the three-body scattering phase of stellar hardening. We stopped the direct $N$-body simulations at a point when the semi-major axis of the BH orbit was much smaller than $a_{\rm h}$. The subsequent evolution of the binary was computed semi-analytically by incorporating combined effects of BH hardening caused by stellar encounters (estimated from $s$; see Table~\ref{Table2}) and energy loss by GW emission. The latter is approximated using the expressions of energy loss by an isolated BH system reported in \citet{Peters_Mathews_1963}. We assume a constant value of eccentricity for our estimates at the time when we stop our simulations. However, as scattering experiments \citep{Sesana_et_al_2011} and numerical simulations \citep{Khan_et_al_2012b,Khan_et_al_2018} show and so does the trend in the simulations presented in the current study, the eccentricity grows in the three-body scattering phase until the onset of strong GW emission, which then circularises the BH binary. Hence, our estimated coalescence time $t_{\rm coal}$ in Table~\ref{Table2} can be shorter, especially for Runs~B and D, which have $e$ values approaching unity [$t_{\rm coal,GW} \sim (1-e^2)^{3.5}$].

We find that, in all four cases, the BHs coalesce in a time much shorter than the Hubble time, within 1--2~Gyr from the beginning of the SPH simulations (when the separation is $\sim$0.1~Mpc) and well within 0.5~Gyr from the formation of a hard BH binary (Figure~\ref{fig:seplucio1}), regardless of the values of remnant triaxiality, BH binary eccentricity, and central stellar density.

The triaxiality of the merger remnant (Figure~\ref{fig:triax}) remains high in general for both the dark matter and stellar distributions. In fact, even a slightly non-spherical (stellar) remnant (as in Run~C) is enough to accommodate BH binary coalescence in less than 0.5 Gyr after its formation. This is consistent with the recent results by \citet{Bortolas_et_al_2018}.

The eccentricity of the orbits (Figure~\ref{fig:eccbbh}) is higher for counter-rotating binaries than in co-rotating binaries (cf. Figs~\ref{fig:multimomgal} and \ref{fig:multimombh}), consistent with results by \citet{Sesana_et_al_2011} and \citet{HolleyBockelmann_Khan_2015}. Again, the BHs coalesce regardless of the value of $e$. However, for similar values of central stellar density, the run with the lowest values of $e$ takes the longest to BH coalescence (Run~A versus C).

On the other hand, for similar values of eccentricity, higher central stellar density values imply shorter coalescence times (Runs~A, B, and C).

The time-scales we obtain (0.025--0.44~Gyr from the beginning of the direct $N$-body simulations) are significantly longer on average than what found in \citet{Khan_et_al_2016}, where they obtain $\sim$10~Myr. This was expected, since in \citet{Khan_et_al_2016} they simulated massive galaxies, with much higher central densities than in our work: at the influence radius, our densities are of the order of 3--$14 \times 10^{11}$~M$_{\odot}$~kpc$^{-3}$, whereas the same value in \citet{Khan_et_al_2016} is $\sim$ $3 \times 10^{13}$~M$_{\odot}$~kpc$^{-3}$. Our relatively low densities are typical of late-type galaxies, and are consistent with what found in cosmological simulations \citep[see, e.g.][]{Bonoli_et_al_2016}. 


\acknowledgments

FMK acknowledges support by Higher Education Commission of Pakistan through NRPU grant 4159. 
PRC acknowledges support by the Tomalla Foundation.
PB acknowledges support by the Chinese Academy of Sciences through the Silk Road Project at NAOC, through the "Qianren" special foreign experts program, and the President's International Fellowship for Visiting Scientists program of CAS, the National Science Foundation of China under grant No. 11673032 and 
also the Strategic Priority Research Program (Pilot B) "Multi-wavelength 
gravitational wave universe" of the Chinese Academy of Sciences (No. 
XDB23040100). For the code development the special GPU accelerated supercomputer Laohu at NAOC has been used and we thank the Center of Information and Computing of NAOC for support. PB acknowledges the support of the Volkswagen Foundation under the Trilateral Partnerships grant No. 90411 and the special support by the NASU under the Main Astronomical Observatory GRID/GPU computing cluster project.
This work benefited from support by the International Space Science Institute, 
Bern, Switzerland,  through its International Team programme ref. no. 393 
"The Evolution of Rich Stellar Populations \& BH Binaries" (2017-18).


\bibliographystyle{aasjournal}
\bibliography{manuscript}

\end{document}